\documentclass[onecolumn]{aastex6}

\usepackage{color,txfonts}
\usepackage{hyperref}
\usepackage{epstopdf}
\usepackage{graphicx}
\usepackage[FIGTOPCAP]{subfigure}

\begin{document}

\title{How Special Is GRB 170817A?}

\author{Chuan Yue$^{1,2}$, Qian Hu$^{1,3}$, Fu-Wen Zhang$^{4}$, Yun-Feng Liang$^{1}$, Zhi-Ping Jin$^{1,5}$,
 Yuan-Chuan Zou$^{6}$, Yi-Zhong Fan$^{1,5}$, and Da-Ming Wei$^{1,5}$}
\affil{
$^1$ {Key Laboratory of dark Matter and Space Astronomy, Purple Mountain Observatory, Chinese Academy of Science, Nanjing 210008, China.}\\
$^2$ {University of Chinese Academy of Sciences, Yuquan Road 19, Beijing 100049, China}\\
$^3$ {Department of Physics and Institute of Theoretical Physics, Nanjing Normal University, Nanjing 210046, China}\\
$^4$ {College of Science, Guilin University of Technology, Guilin 541004, China.}\\
$^5$ {School of Astronomy and Space Science, University of Science and Technology of China, Hefei, Anhui 230026, China.}\\
$^6$ {School of Physics, Huazhong University of Science and Technology, Wuhan 430074, China.}\\
}
\email{fwzhang@pmo.ac.cn (FWZ), yzfan@pmo.ac.cn (YZF) and dmwei@pmo.ac.cn (DMW)}

\begin{abstract}
GRB 170817A is the first short gamma-ray burst (GRB) with direct detection of the gravitational-wave radiation and also the spectroscopically identified macronova emission (i.e., AT 2017gfo). The prompt emission of this burst, however, is underluminous in comparison with the other short GRBs with known redshift. In this work, we examine whether GRB 170817A is indeed unique.
We firstly show that GRB 130603B/macronova may be the on-axis ``analogs" of GRB 170817A/AT 2017gfo, and the extremely dim { but long-lasting} afterglow emission of GRB 170817A may suggest a low number density ($\sim 10^{-5}~{\rm cm^{-3}}$) of its circumburst medium { and a structured outflow}. We then discuss whether GRB 070923, GRB 080121, GRB 090417A, GRB 111005A, and GRB 170817A form a new group of very nearby underluminous GRBs originated from neutron star mergers. If the short events  GRB 070923, GRB 080121, and GRB 090417A are indeed at a redshift of $\sim 0.076,~0.046,~0.088$, respectively, their isotropic energies of the prompt emission are $\sim 10^{47}$ erg and thus comparable to the other two events. The non-detection of optical counterparts of GRB 070923, GRB 080121, GRB 090417A, and GRB 111005A, however, strongly suggests that the macronovae from neutron star mergers are significantly diverse in luminosities or, alternatively, there is the other origin channel (for instance, the white dwarf and black hole mergers). We finally suggest that GW170817/GRB 170817A are likely not alone and similar events will be detected by the upgraded/upcoming gravitational-wave detectors and the electromagnetic monitors.
\end{abstract}


\section{Introduction} \label{sec:intro}
The mergers of close neutron star binaries are
promising sources of gravitational-wave (GW; \citep{1977ApJ...215..311C,LVC2010}) and $r$-process elements \citep{Eichler1989}. Such mergers are also widely believed to be able to generate short gamma-ray bursts \citep[sGRBs; see][]{1993ApJ...413L.101K} and  macronova emission due to the radioactive decay of the formed heavy elements \citep[i.e., the so-called marconova or kilonova; see][]{1998ApJ...507L..59L,2013ApJ...774...25K,2013ApJ...775..113T,2017LRR....20....3M}.
The macronovae/kilonovae are one of the most promising electromagnetic counterparts of binary neutron star mergers \citep{2017LRR....20....3M}. Previously they had been tentatively detected in GRB 130603B \citep{Tanvir2013,Berger2013}, GRB 060614 \citep{2015NatCo...6E7323Y, 2015ApJ...811L..22J} and GRB 050709 \citep{2016NatCo...712898J}. {sGRBs are also attractive electromagnetic counterparts of the neutron star merger events. The sGRB/GW association, however, is usually expected to be only establishable in the 2020s \citep[i.e., in the full sensitivity run of LIGO/Virgo detectors; see][and the references therein]{2015ApJ...809...53C,2016ApJ...827...75L} due to the low detection rate of sGRBs in the local universe \citep{Wanderman2015}. However, such a conclusion was only valid for the energetic sGRBs with a luminosity $\geq 5\times 10^{49}~{\rm erg~s^{-1}}$. As realized recently in \citet{2017arXiv170807008J}, the neutron star mergers detectable for advanced LIGO/Virgo \citep{LVC2010} are very nearby and hence the associated GRBs with a significantly lower luminosity can be measured. The merger-driven GRBs are apparently weak if our line of sight is (slightly) outside of the cone of the relativistic uniform ejecta (i.e., { the off-axis GRBs, as denoted in most literature}) or, alternatively, the outflow is highly structured and the line of sight is outside of the energetic core but still within the ejecta (i.e., { the off-axis structured GRBs}). Clearly, these weak GRBs are helpful in enhancing the GRB/GW association \citep{2017arXiv170807008J}.} Even so, it is still less likely that the first neutron star merger GW event would be accompanied by a sGRB.

On 2017 August 17 at 12:41:06 (GMT), the LIGO/Virgo detectors detected a transient GW signal
that is consistent with a neutron star binary coalescence \citep{Abbott2017b}. Surprisingly, at 12:41:06.47 UT on 17 August 2017,
the $Fermi$ Gamma-ray Burst Monitor (GBM) triggered and located short GRB 170817A \citep{2017ApJ...848L..14G,Abbott2017d}, which is just about 1.7
s after the GW signal and the location also overlaps with the GW event.
{The optical/infrared/ultraviolet follow-up observations \citep[e.g.][]{Coulter2017,Pian2017,Cowperthwaite2017,Kasliwal2017,Tanvir2017,Drout2017,Fong2017,Margutti2017} found a
bright unpolarized source \citep{Covino2017} and the high-quality spectra are
well consistent with the macronova/kilonova model. Initially AT 2017gfo was dominated by the emission from the lanthanide-free outflow component(s)
(i.e., the accretion disk wind and/or the neutrino-driven mass loss of the hypermassive neutron star formed in
the merger), while at late times it was mainly contributed by the emission from the lanthanide-rich region \citep[e.g.][]{Kasen2017,Pian2017,Kasliwal2017,Cowperthwaite2017,Tanvir2017}}. The remarkable GW/GRB/macronova association
is thus well established in the first GW event involving neutron star(s), which directly confirms the long-standing prediction that neutron star mergers can be the source of
GRBs and are a significant site of $r$-process nucleosynthesis \citep{Eichler1989}. Moreover, the GW/GRB/macronova association data impose very tight constraints on the superluminal/subluminal movements of gravitational waves and on the possible violation of weak equivalence principle {\citep[e.g.,][]{Abbott2017d}}, exclude the so-called Dark Matter Emulators and a class of contender models for cosmic acceleration, and rule out the possibility that the progenitors of GRB 170817A are a binary strange star system  \citep{WangH2017}.

In this work, we try to answer one straightforward question: Is GRB 170817A ``alone" or not? For such a purpose we examine two issues. One is whether some bright GRBs could be similar to GRB 170817A if viewed off-axis. The other is whether there are some weak supernova (SN)-less GRBs similar to GRB 170817A.

\section{Are GRB 170817A/AT 2017gfo the Off-axis Analogs of GRB 130603B/Macronova?}\label{sec:GRB130603B}
As already discussed in \citet{2017arXiv170807008J}, usually the GRB ejecta are structured and the off-axis events are likely weaker and more popular than the on-axis bright events. However, for such ejecta, the energy/Lorentz factor distributions are functions of the viewing angle, which are unclear so far. Then in Sec.\ref{sec:II-1}, we simply adopt the uniform jet model and discuss the off-axis scenario (note that in some literature people call it ``off-beam").  While in Sec.\ref{sec:II-2}, we briefly discuss the simplest off-axis structured jet model (i.e., the two-component jet model) for the afterglow emission.

\subsection{GRB 170817A/AT 2017gfo as the Off-axis Analogs of GRB 130603B/Macronvoa}\label{sec:II-1}
In \citet{WangH2017} we have compared GRB 170817A, the first short event unambiguously powered by neutron star mergers, with other sGRBs with the well-measured spectra and redshifts, and found out that GRB 170817A is the weakest sGRB detected so far and its isotropic equivalent energy ($E_{\rm iso}$) and luminosity ($L_\gamma$) of the prompt emission are at least tens of times lower than the majority recorded before. However, its spectral peak energy, $E_{\rm p,rest}=187 \pm 63$ keV \citep{2017ApJ...848L..14G}, is comparable to quite a few sGRBs, where the subscript `rest' represents the property in the rest frame. Therefore, GRB 170817A does not follow
the regular $E_{\rm p,rest}-E_{\rm iso}$ and $E_{\rm p,rest}-L_\gamma$ correlations presented, for example, in \citet{ZhangFW2012,Tsutsui2013,DAvanzo2014}. One possibility is that GRB 170817A is an off-axis event, which satisfies the following relations
\begin{equation}
E_{\rm iso}\approx a_0^{-3}E_{\rm iso,int},~~L_{\rm \gamma}\approx a_0^{-4}L_{\rm \gamma,int},~~T_{90}\sim a_0 T_{\rm 90,int}, ~{\rm and}~~E_{\rm p}\sim a_0^{-1} E_{\rm p,int}
\end{equation}
where $a_0=1+(\eta \Delta\theta)^{2}$, $\eta$ is the initial Lorentz factor of the GRB outflow, $\Delta \theta=\theta_{\rm v}-\theta_{\rm j}\geq 0$, where $\theta_{\rm v}$ is the viewing angle and $\theta_{\rm j}$ is the half-opening angle, $T_{90}$ is the duration of the GRB, and the subscript ``int" represents the parameters of the event if viewed on-beam \citep[e.g.][]{Yamazaki2002}. Since $E_{\rm p}$ declines much shallower than either $E_{\rm iso}$ or $L_{\gamma}$, the regular $E_{\rm p,rest}-E_{\rm iso}$ and $E_{\rm p,rest}-L_\gamma$ correlations are thus violated. The main purpose of this subsection is to examine whether there are any proper candidates that can serve as the ``on-axis" analogs of GRB 170817A.

\begin{figure}[ht!]
\figurenum{1}\label{fig:comparision}
\centering
\includegraphics[angle=0,scale=0.45]{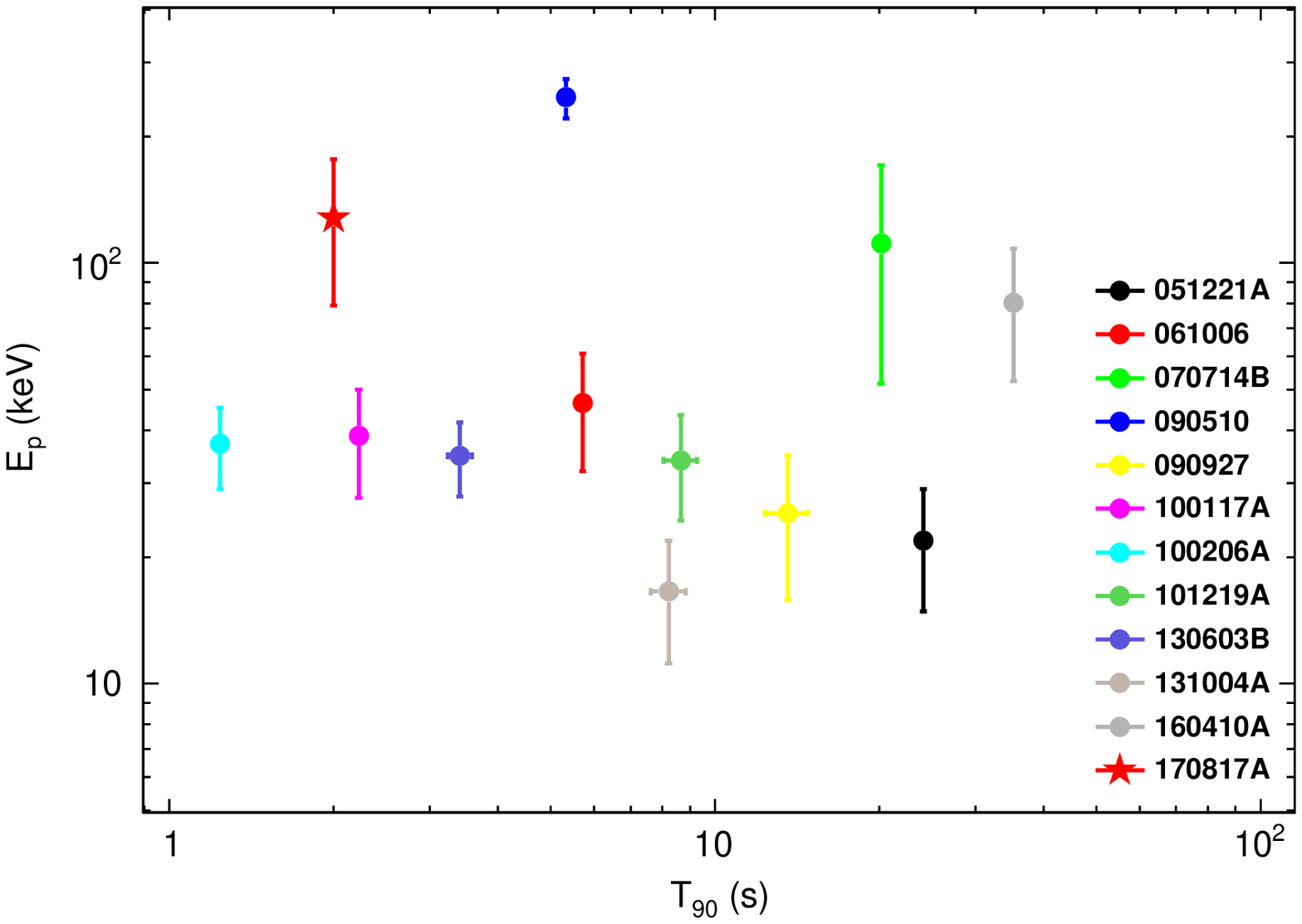}
\includegraphics[angle=0,scale=0.65]{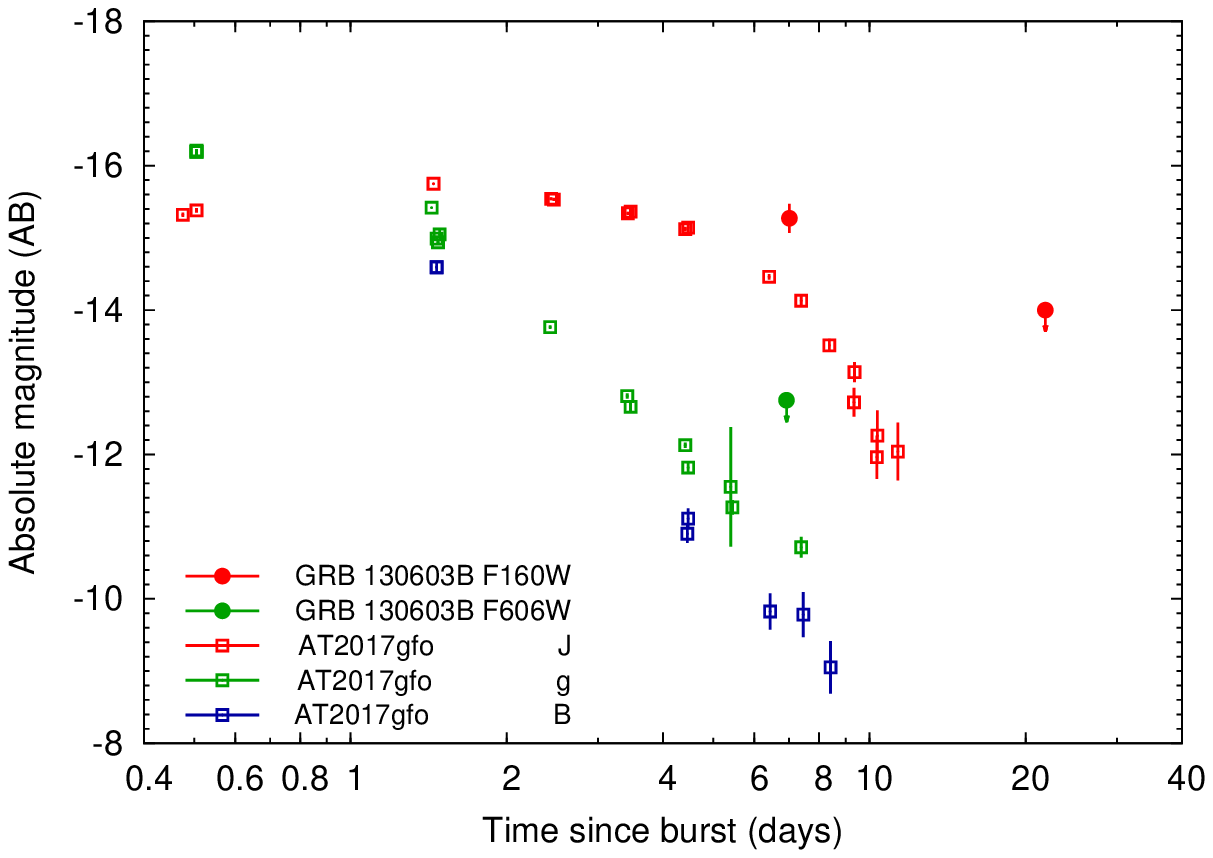}
\caption{Left panel shows the ``extended" durations but ``lowered" spectral peak energies, both by a factor of $a_0\equiv (L_{\rm \gamma,int}/10^{47}~{\rm erg})^{1/4}$, of the bright sGRBs with an observed $L_{\rm \gamma} >10^{51}~{\rm erg~s^{-1}}$ (i.e., these bursts are assumed to be viewed ``off-axis"). The pentagram represents the data of GRB 170817A. One can see that a few ``off-viewed" events (for example, GRB 100206A, GRB 100117A, and GRB 130603B) are similar to GRB 170817A. {The ``intrinsic" data are either taken
from \citet{ZhangFW2012}, \citet{DAvanzo2014}, \citet{Gruber2014} and \citet{2017ApJ...848L..14G} or analyzed in this work.} The right panel is the $J,~g,~B$ band light curves of AT 2017gfo, in comparison to the F160W/F606W-band observations of GRB 130603B (with corrections from $z=0.356$ to $z=0.0097$). {AT 2017gfo was observed by many facilities \citep{Abbott2017c}. Among them, we have chosen the world's largest optical telescopes with relatively rapid and dense observations, including that by VLT \citep{Pian2017,Tanvir2017}, Gemini-South \citep{Kasliwal2017,Cowperthwaite2017}, Magellan \citep{Drout2017}, DECam \citep{Cowperthwaite2017} and VISTA \citep{Tanvir2017}. The optical data of GRB 130603B are from \citet{Tanvir2013}, but see also \citet{Berger2013}}.}
\hfill
\end{figure}

We have collected the bright sGRBs (i.e., the observed $L_{\rm \gamma} >10^{51}~{\rm erg~s^{-1}}$, which is denoted as $L_{\rm \gamma,int}$) detected so far \citep{ZhangFW2012,DAvanzo2014,WangH2017}. Then, we calculate $a_0=(L_{\rm \gamma,int}/10^{47}~{\rm erg~s^{-1}})^{1/4}$ (i.e, we take the isotropic luminosity of GRB 170817A, which is $\sim 10^{47}~{\rm erg~s^{-1}}$, \citep{2017ApJ...848L..14G}, as a fiducial value) and then get the $T_{90}$ and $E_{\rm p}$ of these ``off-axis" events. The results are shown in the left panel of Fig.\ref{fig:comparision}. For comparison, GRB 170817A is also presented (but without any change of the parameters since it is speculated to be an off-axis event). {As expected, almost all of these ``off-axis" events peak in soft gamma-rays with a duration that should be classified into the long-duration group (except that the prompt emission pulses are so short that with the enhancement by a factor of $a_0$ it is still shorter than the central engine activity interval), indicating that}  {some long duration GRBs that
have no associated supernovae} (also known as the SN-less long GRBs or the hybrid GRBs) might be actually the ``off-axis" sGRBs and thus associated with strong GW radiation \citep{WangYZ2017}. Nevertheless, a few events (for example, GRB 100206A, GRB 100117A, and GRB 130603B if viewed off-axis) have parameters similar to GRB 170817A. Note that GRB 130603B is a short event with a macronova signal \citep{Tanvir2013,Berger2013}. In the double neutron star merger model, the late time macronova emission is largely isotropic, \footnote{
A few days after the burst the macronova emission should be mainly from the lanthanide-rich outflow (i.e., the dynamical ejecta). As found in binary neutron star merger simulations, such outflow components are largely isotropic \citep[see, e.g. Fig.1 of][]{Hotokezaka2013}. At early times the macronova emission however may depend on the viewing angle sensitively since in some directions the outflow may be lanthanide-free that yields bluer/brighter radiation \citep{2017LRR....20....3M}.}, which is remarkably different from the prompt $X$-/$\gamma$-ray emission. Though such a macronova signature has just one data point, it indeed reasonably matches the $J$-band emission of AT 2017gfo, strengthening the potential ``connection" between GRB 170817A and GRB 130603B (see the right panel of Fig.\ref{fig:comparision}).

\subsection{Low Circumburst medium of GRB 170817A?}\label{sec:II-2}
{The prompt emission and the macronova emission are respectively governed by the relativistic and sub-relativistic outflows launched during the merger of the binary compact objects, which are independent of the density of the circumburst medium \citep[e.g.,][]{1998ApJ...507L..59L,2013ApJ...774...25K,2013ApJ...775..113T,2017LRR....20....3M,Fernandez2017}.}
The afterglow emission is instead generated by the electrons accelerated in the blast wave expanding into the medium.
Therefore, its flux depends on the medium density ($n$) sensitively, except that the observer's frequency is above both the cooling frequency and the typical synchrotron radiation frequency of the electrons, for which the flux is proportional to $n^{0}$  \citep{Sari1998}. If GRB 170817A is an { off-axis analog} of GRB 130603B but within the medium with a significantly lower $n$ (note that for GRB 130603B the afterglow modeling fit suggests an $n \sim 0.15~{\rm cm^{-3}}$ \citep{Fan2013ApJL}, which is higher than that for typical sGRBs; \citep{Fong2015}), the afterglow emission would be further suppressed (in the off-axis jet scenario, the early time afterglow should rise rather than decay until the ejecta has got decelerated to a Lorentz factor that is already smaller than $1/\theta_{\rm v}$, where $\theta_{\rm v}$ is the observer's viewing angle \citep{Granot2002,Wei2003}).

\begin{figure}[ht!]
\figurenum{2}\label{fig:afterglow}
\centering
\includegraphics[angle=0,scale=0.55]{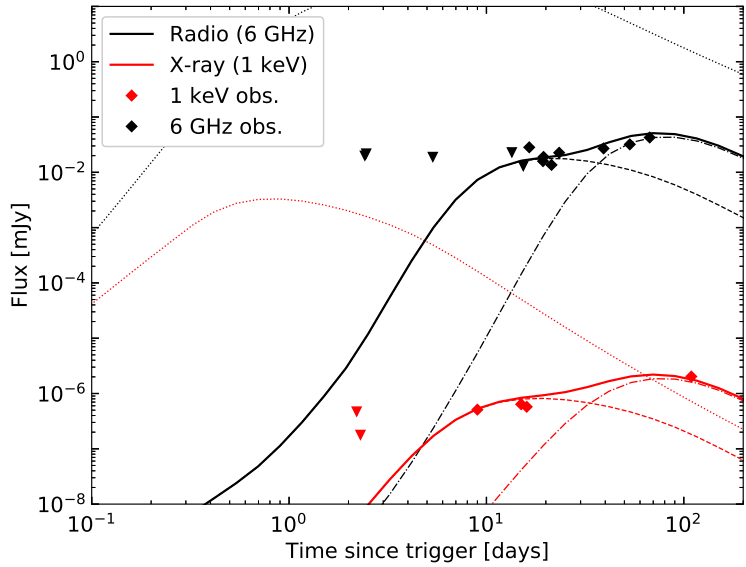}
\caption{Dotted lines represent the afterglow emission of GRB 130603B if observed at the viewing angle of $\theta_{\rm v}\sim 0.185$ rad and shifted to the redshift of $z=0.0097$.  To better match the first month of afterglow data of GW170817/GRB 170817A, we have adjusted $n$, $p$ and $\epsilon_{\rm B}$ to be $2.5\times10^{-5}~{\rm cm^{-3}}$, $2.15$ and $3.0\times10^{-4}$, respectively (see the dashed lines). { An additional energetic core component, with the isotropic equivalent kinetic energy $\sim 1.3\times 10^{52}\,{\rm erg}$ and the half-opening angle $\sim0.5\,\theta_{\rm j}$, is  added into the model to match the late-time radio and X-ray data (see the dotted-dashed lines).} The solid lines represent the total afterglow emission (the X-ray and radio emission are in red and black, respectively). { The X-ray data points (flux at 1 keV) of GW170817/GRB 170817A are adopted from \citet{Margutti2017}, \citet{Troja2017}, \citet{Haggard2017} and \citet{Ruan2017}.} The XRT upper limits, well above the red solid line, are not shown. The radio data points are from \citet{Alexander2017}, \citet{Hallinan2017} and \citet{Mooley2017}.}
\hfill
\end{figure}

The afterglow emission of GRB 130603B {\citep[e.g.,][]{Tanvir2013,Berger2013,deU2014,Fong2014}} have been reasonably fitted \citep{Fan2013ApJL}. We adopt the same set of physical parameters (including the energy injection process assumed in \citet{Fan2013ApJL}) to estimate the off-axis afterglow radiation. The physical parameters include $\epsilon_{\rm e}=0.15$ and $\epsilon_{\rm B}= 0.03$ (i.e., the fractions of shock energy given to electrons and magnetic field), $\theta_{\rm j}=0.085$, the initial kinetic energy of the GRB ejecta, is adopted to be $2\times 10^{50}$ erg; and the power-law spectral index of electrons $p=2.3$.  The energy injection into the blast wave has been assumed to be $\approx 1.6\times 10^{48}~{\rm erg~s^{-1}}(1+t/680~{\rm s})^{-1}$ for $t\leq 680$ s, otherwise it is zero (i.e., after the energy injection, the total isotropic equivalent kinetic energy is of $\sim 1.2\times 10^{51}$ erg). The initial Lorentz factor ($\Gamma_0$) is unknown. The empirical correlation between the prompt emission luminosity and the bulk Lorentz factor \citep{Lv2012,Liang2010,Fan2012ApJL} suggests a typical value  $\Gamma_0\sim 100$, while a value lowered by a factor of a few is allowed. In this work, we take $\Gamma_0\approx 50$ and then estimate $\theta_{\rm v}$ as $\sim \theta_{\rm j}+(L_{\rm \gamma,int}/10^{47})^{1/8}\Gamma_0^{-1}\sim 0.185$ rad. With such a set of parameters, we calculated the afterglow emission with the same code adopted by \citet{Fan2013ApJL} but move the source to a redshift of 0.0097.
%The results are shown in Fig.\ref{fig:afterglow}.
{The resulting afterglow emission is far brighter than that observed in GRB 170817 (see the dotted lines in Fig.\ref{fig:afterglow}). In order to match the afterglow data (or upper limits) of GRB 170817A collected in the first month after the burst, the physical parameters are adjusted to be $n\sim 2.5\times10^{-5}~{\rm cm^{-3}}$, $p\sim2.15$, and $\epsilon_{\rm B}\sim3.0\times10^{-4}$ (see the solid lines in Fig.\ref{fig:afterglow}). { To reasonably address the latest (i.e., $t\sim 100$ days) afterglow data,  an additional energetic/narrow core, with the isotropic equivalent kinetic energy $\sim 1.3\times 10^{52}\,{\rm erg}$ and the half-opening angle $\sim0.5\,\theta_{\rm j}$, is likely needed \citep[see][and the references therein for more discussion on the structured outflows of short GRBs]{2017arXiv170807008J}. Such an energetic jet component is expected to have a bulk Lorentz factor of hundreds, hence its contribution to the prompt emission can be ignored. Other possible interpretations can be found in \citep{Mooley2017,Pooley2017}.} Finally, we would like to point out that both $n$ and $\epsilon_{\rm B}$ are in the low end of the distribution of the physical parameters for short GRB afterglows \citep[see][]{Fong2015}, implying brighter afterglow emission in some other neutron star merger events.}

\section{A new group of Very Nearby Underluminous Merger-driven events?}
%\subsection{The prompt emission properties of GRB 170817A, GRB 080121 and GRB 111005A}

In the previous section, we have shown that the prompt emission of some sGRBs, if viewed off-axis, could have an $E_{\rm p}\sim 100$ keV and lasts a few seconds, which is similar to
GRB 170817A. One interesting event is GRB 130603B, a burst with a macronova signal. The afterglow emission of the off-axis ejecta may emerge at late times. If 
GRB 170817A is indeed related to a typical merger-driven GRB, similar events should have already been recorded by {\it Swift} {or other monitors \citep[see][for an early effort with the BATSE sample]{Tanvir2005}}. This can be estimated as the follows. The successful detection of a binary neutron star merger event by advanced LIGO/Virgo in the O2 Run yields a local neutron star merger rate of $R_{\rm nsm} \sim 1540^{+3200}_{-1220}~{\rm Gpc^{-3}~yr^{-1}}$ \citep{Abbott2017b}. Intriguingly, the local bright sGRB data independently suggest a merger rate of $\sim 300-1000~{\rm Gpc^{-3}~yr^{-1}}$ \citep{2017arXiv170807008J}. Therefore, within the sensitivity horizon (i.e., $z\sim 0.1$ for double neutron star mergers) of advanced LIGO/Virgo for GW/GRB association events, {\it Swift}/BAT should have already detected $\sim 1~({\cal R}_{\rm nsm}/10^{3}~{\rm Gpc^{-3}~yr^{-1}})(\theta_{\rm j}/0.1)^{2}$ bright sGRBs and $\sim 10~({\cal R}_{\rm nsm}/10^{3}~{\rm Gpc^{-3}~yr^{-1}})(\theta_{\rm j}/0.4)^{2}$ underluminous sGRBs. Here, we focus on the {\it Swift} bursts since the BAT detector is more sensitive than $Fermi$-GBM, $INTEGRAL$ and Konus-$Wind$, at lower energy band. {\it Swift} BAT can locate the very weak short bursts accurately (i.e., with a typical error of a few arcminutes), guide the followup observations and then identify host galaxies \citep{Gehrels2004}.

 \subsection{Very-nearby underluminous merger driven events: a few additional candidates}

\begin{figure}[ht!]
\figurenum{3}\label{fig:group}
\centering
\includegraphics[angle=0,scale=0.6]{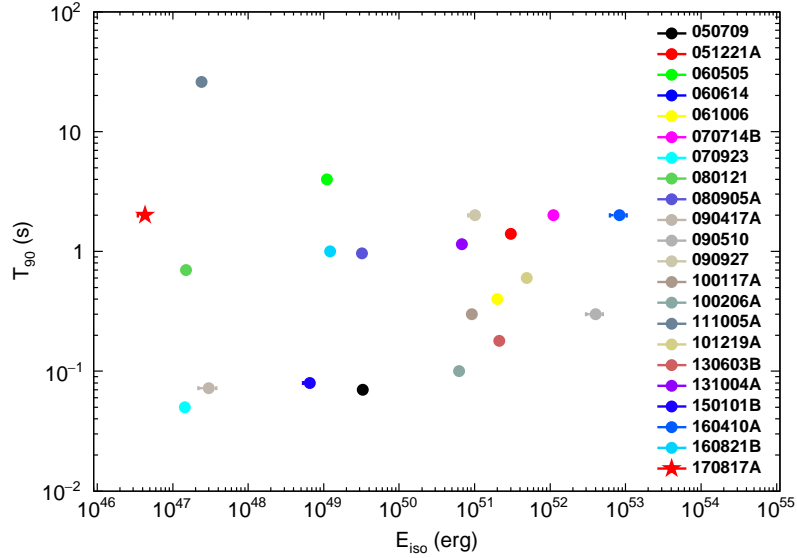}
\caption{Distribution of $T_{90}$ and $E_{\rm iso}$ of sGRBs and {a few long-duration GRBs that
have no associated supernovae} (including GRB 060505, GRB 060614, and GRB 111005A). There are five underluminous events, including GRB 170817A, GRB 111005A, and three bursts with tentative redshifts, e.g., GRB 070923, GRB 080121, and GRB 090417 ($z=0.076,~0.046,~0.088$ are assumed, respectively). These events are at least a few tens times less energetic than the others. Such a fact suggests that GRB 170817A is not alone and there is likely a group of subluminous events that have a high rate density.}
\hfill
\end{figure}

Figure \ref{fig:group} shows the distribution of $T_{90}$ and $E_{\rm iso}$ {(calculated in the rest frame in the energy range $1-10,000$ keV)} of sGRBs and long-short GRBs  (including GRB 060505, GRB 060614, and GRB 111005A) that also could be from neutron star mergers. {The data are taken from \citet{ZhangFW2012}, \citet{DAvanzo2014}, \citet{Gruber2014} and \citet{LiYe2016} and some are analyzed in this work.}  In the left side of the plot we find a few additional underluminous events (or candidates), including GRB 111005A \citep[$z=0.0133$; see][]{Michal2017,Tanga2017}, GRB 070923, GRB 080121, and GRB 090417A, which are at least a few tens of times smaller than the others.  {The caution is that for GRB 070923, GRB 080121, and GRB 090417A, the redshifts are not secure, simply because of the lack of the quick follow-up observations (prompt {\it Swift}/XRT observations were not possible for these events due to the sun constraint).  Their tentative redshifts $\sim (0.076,~0.046,~0.088)$ are based on the possible association of the bursts with nearby galaxies \citep{Fox2007,Perley2008,Fox2009}.} %Nevertheless, we are keen on whether there are any candidates for a new group of underluminous merger-driven events}
Though the galaxy NGC 3313 ($z=0.0124$) in the direction of GRB 150906B (a very hard short GRB) has been proven to be just a coincidence \citep{Abbott2017a}, the possibility of having the whole sample (now consisting of four candidates) solely by chance is very small. Moreover, the presence of GRB 111005A, a long-short event at $z=0.0113$ \citep{Michal2017,Tanga2017},
strongly favors the ``emergence" of a group of underluminous merger-driven GRBs.
In \citet{WangYZ2017} we have estimated the rate density of a GRB 111005A-like event to be ${\cal R}=58^{+219}_{-38} ~{\rm Gpc^{-3}~yr^{-1}}$ (using the Uniform prior, the errors are reported in $90\%$ confidence level) or $29^{+199}_{-18}~{\rm Gpc^{-3}~yr^{-1}}$ (using the Poisson Jeffreys prior). Note that the jet-opening angle correction has not been taken into account and hence the intrinsic rate density could be much higher. The rate density of GRB 170817A is difficult to estimate since usually the $Fermi$-GBM GRBs have very large localization error and it is rather challenging to carry out intense follow-up observations and then establish their connection with the ``nearby" galaxies.  Assuming that GRB 170817A was observed at a viewing angle of $\sim 0.4$ rad, the corresponding GRB 170817A-like event rate is $\sim 16-640~{\rm Gpc^{-3}~yr^{-1}}$, well consistent with ${\cal R}$. Such a fact further strengthens the possible connection between GRB 170817A and GRB 111005A.

%\begin{table}
%\begin{center}
%\label{tab:SGRBjet}
%\title{}Table 1. Some emission properties of five underluminous sGRBs. \\
%\begin{tabular}{cccccc}
%\hline
%\hline
%GRB & $z$	& $E_{{\rm iso}}$	& $T_{90}$ 	& ${\rm macronova}$	& References \\
% 	& 	&	($10^{47}$erg)	& (s)	& 	& \\
%\hline
%GRB 170817A	& 0.0097	& 0.43	& 2	& bright & (1) \\
%GRB 070923  & 0.076 (?) & 1.45  & 0.05  &  no evidence  & (2)            \\
%GRB 080121	& 0.046 (?)	&  1.5	& 0.7   & no evidence  & (3)              \\
%GRB 090417A & 0.088 (?) & 3.0   &0.072  & no evidence  &(4)                \\
%GRB 111005A	& 0.0133	& 2.4	& 26 & Stringent limit	& (5,6)         \\

%\hline
%\end{tabular}
%\end{center}
%Maybe it is better to also list the luminosities, which may be more useful.\\
%References: (1) \citet{Goldstein2017}; (2) \citet{Fox2007}; (3) \citet{Troja2008}; (4)\citet{Fox2009}; (5)\citet{Michal2017};(6) \citet{Tanga2017}.
%\end{table}

\subsection{Diverse macronova emission}
\begin{figure}[ht!]
\figurenum{4}\label{fig:macronova}
\centering
\includegraphics[angle=0,scale=0.70]{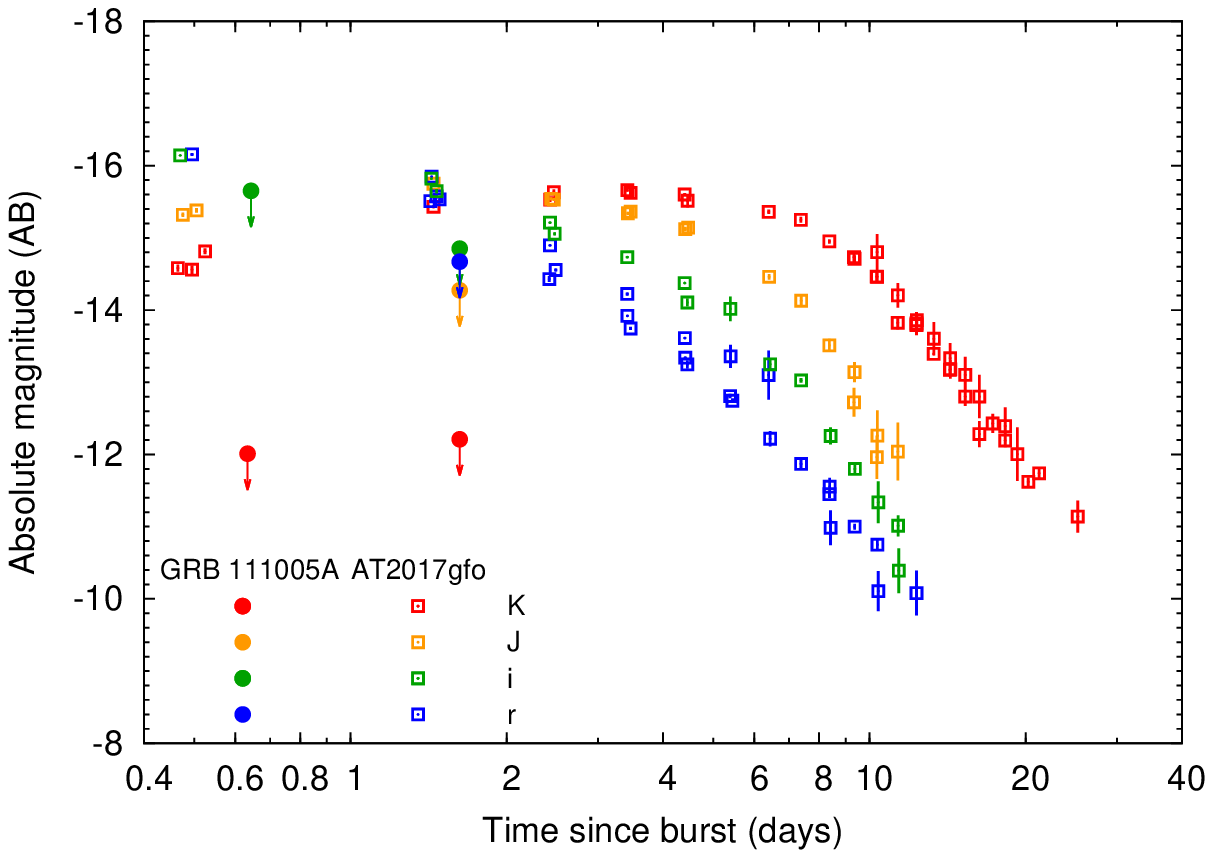}
\includegraphics[angle=0,scale=0.70]{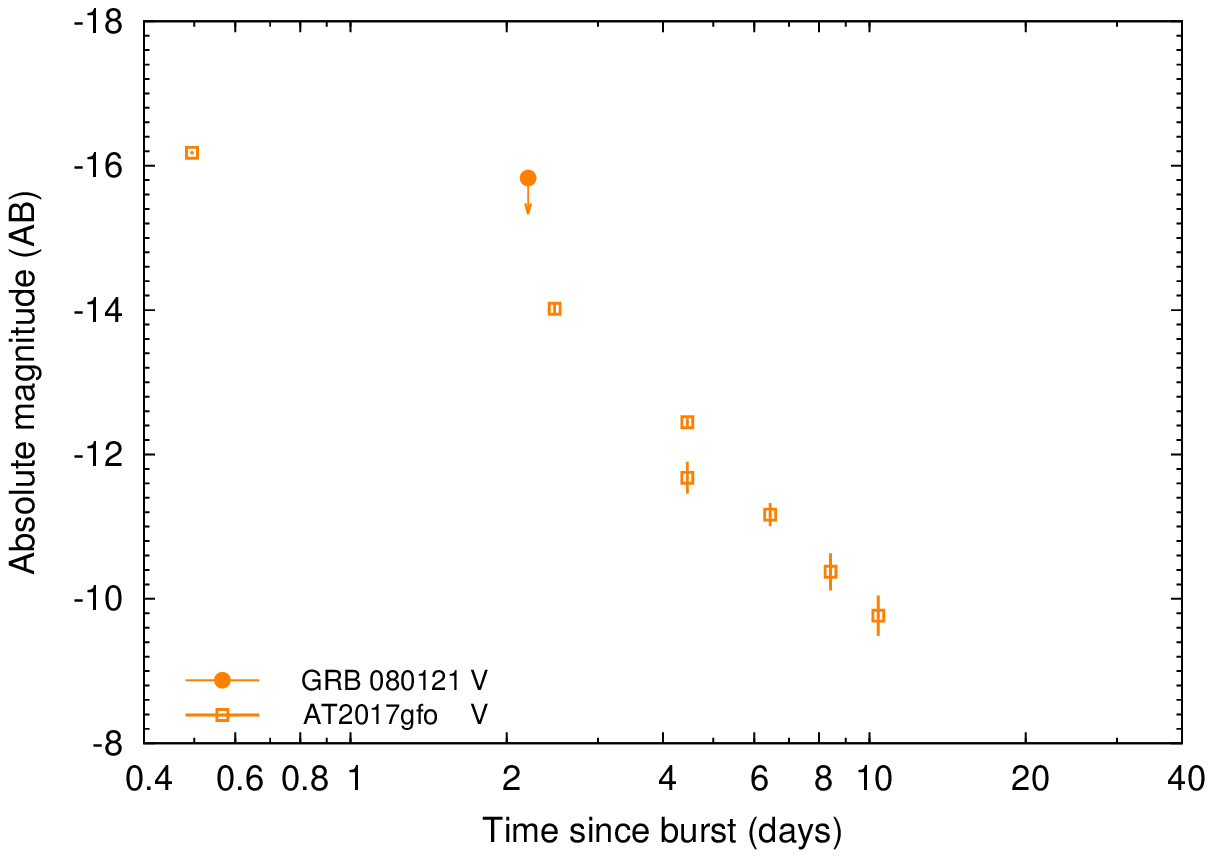}

\caption{AT 2017gfo in comparison to the upper limits of GRB 111005A (left panel) and GRB 080121 (right panel). The data of AT 2017gfo {are adopted from the same set of references quoted  in Fig.1}. The {upper limits} of GRB 111005A and GRB 080121  are taken from \citet{Tanga2017} and \citet{Troja2008}. Note that for GRB 111005A the extinction corrections on the upper limits have been taken into account.}
\hfill
\end{figure}

GRB 170817A hosts the first spectroscopically identified  macronova AT 2017gfo. While for GRB 111005A, GRB 070923, GRB 080121, and GRB 090417A, no reliable optical/infrared counterparts have been detected. There are just some upper limits, and we have compared to AT 2017gfo (Note that all of these upper limits have been properly corrected to the redshift of GRB 170817A. For the observation data of GRB 111005A, the extinction correction with $A_{\rm v}=2$ mag also has been addressed). As shown in Fig.\ref{fig:macronova}, the macronova associated with GRB 111005A should be dimmer than AT 2017gfo by 2-3 mag in the $K$ band, which might imply that the macronova emission from neutron star mergers are diverse {(see also \citet{Gompertz2017} and \citet{Fong2017} for similar conclusions, though GRB 111005A was not included in these analyses).}  For GRB 080121, the constraint is much looser due to its relatively high redshift, while for GRB 070923 and GRB 090417A there were no optical observations. The diversity of the macronova radiation may be caused by the different physical properties of the merger system as well as our lines of sight. In view of the high neutron star merger rate, tens of macronovae triggered by the gravitational-wave detection are expected to be annually detected in 2020s, and our speculation can be directly tested.

\section{Summary} \label{sec:discussion}
GRB 170817A is the first short GRB with direct detection of the gravitational-wave radiation and also the spectroscopically identified macronova emission (i.e., AT 2017gfo). GW170817 is further distinguished as the first gravitational wave signal from the double neutron star merger. Though the association of sGRB, GW, and macronova is widely expected,
the firm establishment of the GW/GRB/macronova association in the first double neutron star merger event recorded by Advanced LIGO/Virgo is completely a surprise to the community.
GRB 170817A is the weakest short GRB identified so far, and one straightforward speculation is that the dim prompt emission is due to our large viewing angle to the center of the relativistic ejecta {\citep[see, e.g.][]{Margutti2017,Troja2017}.} If correct, it implies that there are some bright sGRBs that can appear as GRB 170817A if viewed off-axis and there are more sub-luminous events in the {\it Swift} data. In this work, we have examined these two possibilities. We found out that a few bright sGRBs, if viewed sideways, may resemble GRB 170817A. In particular, GRB 130603B/macronova might be ``on-axis" analogs of GRB 170817A/AT 2017gfo. The extremely dim afterglow emission of GRB 170817A, though as close as $z=0.0097$, likely suggests a rather low number density of its circumburst medium. We also found out that there may be a new group of very nearby underluminous sGRBs originated from neutron star mergers. Particularly, if the short events GRB 070923, GRB 080121, and GRB 090417A are indeed at a redshifts of $\sim 0.076,~0.046,~0.088$, respectively, their isotropic energies of the prompt emission are $\sim 10^{47}$ erg and thus comparable to the GRB 170817A and GRB 111005A. The non-detection of optical counterparts of GRB 080121 and GRB 111005A, however, strongly suggests that the macronovae from neutron star mergers are significantly diverse in luminosities {\citep[see e.g.][]{Gompertz2017,Fong2017}}  or, alternatively, there is another origin channel (for instance, the white dwarf and black hole mergers). We finally conclude that GW170817/GRB 170817A are not alone and similar events will be jointly detected by the upgraded/upcoming gravitational-wave detectors and the electromagnetic monitors.

\section*{Acknowledgments}
{We thank the anonymous referee for helpful suggestions.} This work was supported in part by 973 Programme of China (No. 2014CB845800), by NSFC under grants 11525313 (the National Natural Fund for Distinguished Young Scholars), 11433009, 11763003, and 11773078, by the Chinese Academy of Sciences via the Strategic Priority Research Program (No. XDB23040000), Key Research Program of Frontier Sciences (No. QYZDJ-SSW-SYS024), and the External Cooperation Program of BIC (No. 114332KYSB20160007).

\clearpage


\begin{thebibliography}{}
\bibitem[Abadie et al. (2010)]{LVC2010} Abadie, J., Abadie, J., Abbott, B. P., et al. 2010, CQGra, 27, 173001
\bibitem[Abbott et al.(2016)]{2016PhRvX...6d1015A} Abbott, B.~P., Abbott, R., Abbott, T.~D., et al.\ 2016, Physical Review X, 6, 041015
\bibitem[Abbott et al. (2017a)]{Abbott2017a} Abbott, B.~P., Abbott, R., Abbott, T.~D., et al. 2017a, \apj, 841, 89
\bibitem[Abbott et al. (2017b)]{Abbott2017b} Abbott, B.~P., Abbott, R., Abbott, T.~D., et al. 2017b, Phys. Rev. Lett., 119, 161101
\bibitem[Abbott et al.(2017c)]{Abbott2017c} Abbott, B.~P., Abbott, R., Abbott, T.~D., et al.\ 2017c, \apjl, 848, L12
\bibitem[Abbott et al.(2017d)]{Abbott2017d} Abbott, B.~P., Abbott, R., Abbott, T.~D., et al.\ 2017c, \apjl, 848, L13
\bibitem[Alexander et al.(2017)]{Alexander2017} Alexander, K.~D., Berger, E., Fong, W., et al.\ 2017, \apjl, 848, L21
\bibitem[Berger et al.(2013)]{Berger2013} Berger, E., Fong, W., \& Chornock, R.\ 2013, \apjl, 774, L23
\bibitem[Clark \& Eardley (1977)]{1977ApJ...215..311C} Clark, J.~P.~A., \& Eardley, D.~M.\ 1977, \apj, 215, 311
\bibitem[Clark et al.(2015)]{2015ApJ...809...53C} Clark, J., Evans, H., Fairhurst, S., et al.\ 2015, \apj, 809, 53
\bibitem[Coulter et al. (2017)]{Coulter2017} Coulter, D., et al., Gamma Ray Coordinates Network Circular 21567 (2017).
\bibitem[Covino et al.(2017)]{Covino2017} Covino, S., et al. 2017, Nature Astronomy, 1£¬ 791
\bibitem[Cowperthwaite et al.(2017)]{Cowperthwaite2017} Cowperthwaite, P. S., Berger, E., Villar, V. A. et al. 2017, ApJL, 848, L17
\bibitem[D'Avanzo et al.(2014)]{DAvanzo2014} D'Avanzo, P., Salvaterra, R., Bernardini, M.~G., et al.\ 2014, \mnras, 442, 2342
\bibitem[Drout et al.(2017)]{Drout2017} Drout, M. R., Piro, A. L., Shappee, B. J. et al. 2017, arXiv:1710.05443
\bibitem[de Ugarte Postigo et al.(2014)]{deU2014} de Ugarte Postigo, A., et al. 2014, A\&A,  563, 62
\bibitem[Eichler et al. (1989)]{Eichler1989} Eichler, D., Livio, M., Piran, T., \& Schramm, D.~N.\ 1989, \nat, 340, 126
\bibitem[Fan et al. (2012)]{Fan2012ApJL} Fan, Y. Z., et al. 2012, \apjl, 755, L6
\bibitem[Fan et al. (2013)]{Fan2013ApJL} Fan, Y. Z., et al. 2013, \apjl, 779, L25
\bibitem[Fong et al. (2014)]{Fong2014} Fong, W. F., et al. 2014, \apj, 780, 118
\bibitem[Fong et al. (2015)]{Fong2015} Fong, W. F., et al. 2015, \apj, 815, 102
\bibitem[Fong et al. (2017)]{Fong2017} Fong, W. F., et al. 2017, \apj, 848, L23
\bibitem[Fox (2009)]{Fox2009} Fox, D. B., 2009, GCN Circ. 9134
\bibitem[Fox \& Ofek (2007)]{Fox2007} Fox, D. B., \& Ofek, E. 2007, GCN Circ. 6819
\bibitem[Fern\'{a}ndez et al. (2017)]{Fernandez2017} Fern\'{a}ndez, R., et al. 2017,
 Classical and Quantum Gravity, 34, 154001
\bibitem[Gehrels et al.(2004)]{Gehrels2004} Gehrels, N. et al., 2004, \apj, 611, 1005-1020
\bibitem[Goldstein et al. (2017)]{2017ApJ...848L..14G} Goldstein, A., Veres, P., Burns, E., et al.\ 2017, \apjl, 848, L14
\bibitem[Gompertz  et al. (2017)]{Gompertz2017} Gompertz, B. P. et al. 2017, arXiv:1710.05442
\bibitem[Granot et al. (2002)]{Granot2002} Granot, J., Panaitescu, A., Kumar, P., \& Woosley, S. E., 2002, ApJ, 570, L61
\bibitem[Gruber et al. (2014)]{Gruber2014} Gruber, D., Goldstein, A., Weller von Ahlefeld, V., et al. 2014, ApJS, 211, 12
\bibitem[Haggard et al.(2017)]{Haggard2017} Haggard, D., Nynka, M., Ruan, J. et al. 2017, ApJL, 848, L25
\bibitem[Hallinan et al. (2017)]{Hallinan2017} Hallinan, G., et al. 2017, Science, doi:10.1126/science.aap9855
\bibitem[Hotokezaka et al. (2013)]{Hotokezaka2013} Hotokezaka, K. et al. 2013, \apjl, 778, L16
\bibitem[Jin et al.(2015)]{2015ApJ...811L..22J} Jin, Z.-P., Li, X., Cano, Z., et al.\ 2015, \apjl, 811, L22
\bibitem[Jin et al.(2016)]{2016NatCo...712898J} Jin, Z.-P., Hotokezaka, K., Li, X., et al.\ 2016, Nature Communications, 7, 12898
\bibitem[Jin et al.(2017)]{2017arXiv170807008J} Jin, Z.-P., Li, X., Wang, H., et al.\ 2017, arXiv:1708.07008
\bibitem[Kasen et al.(2013)]{2013ApJ...774...25K} Kasen, D., Badnell, N.~R., \& Barnes, J.\ 2013, \apj, 774, 25
\bibitem[Kasen et al.(2017)]{Kasen2017} Kasen, D., Metzger, B., \& Barnes, J.\ 2017, Natur., 551, 80
\bibitem[Kasliwal et al.(2017)]{Kasliwal2017} Kasliwal, M. M., et al. 2017, Science, doi:10.1126/science.aap9455
\bibitem[Kouveliotou et al.(1993)]{1993ApJ...413L.101K} Kouveliotou, C., Meegan, C.~A., Fishman, G.~J., et al.\ 1993, \apjl, 413, L101
\bibitem[Li \& Paczy{\'n}ski(1998)]{1998ApJ...507L..59L} Li, L.-X., \& Paczy{\'n}ski, B.\ 1998, \apjl, 507, L59
\bibitem[Li et al.(2016a)]{2016ApJ...827...75L} Li, X., Hu, Y.-M., Fan, Y.-Z., \& Wei, D.-M.\ 2016, \apj, 827, 75
\bibitem[Li et al.(2016b)]{LiYe2016} Li, Y., Zhang, B., \& L{\"u}, H.-J.\ 2016b, \apjs, 227, 7
\bibitem[Liang et al. (2010)]{Liang2010} Liang, E.-W., Yi, S.-X., Zhang, J., et al. 2010, ApJ, 725, 2209
\bibitem[L\"{u} et al. (2012)]{Lv2012} L\"{u}, J., et al. 2012, \apj, 751, 49
\bibitem[Margutti et al.(2017)]{Margutti2017} Margutti, R., Berger, E., Fong, W. et al. 2017, ApJL, 848, L20
\bibitem[Metzger(2017)]{2017LRR....20....3M} Metzger, B.~D.\ 2017, Living Reviews in Relativity, 20, 3
\bibitem[Michal et al. (2017)]{Michal2017} Michal, J., et al. 2017, arXiv:1610.06928
\bibitem[Mooley et al.(2017)]{Mooley2017} Mooley, K.~P., Nakar, E., Hotokezaka, K., et al.\ 2017, arXiv:1711.11573
\bibitem[Perley et al. (2008)]{Perley2008} Perley, D. A., Foley, R. J., \&  Bloom, J. S., 2008, GCN Circ. 7210
(https://gcn.gsfc.nasa.gov/gcn/gcn3/7210.gcn3)
\bibitem[Pian et al.(2017)]{Pian2017} Pian, E., et al.\ 2017, Nature, 551£¬ 67
\bibitem[Pooley et al. (2017)]{Pooley2017} Pooley, D., Kumar, P. \& Wheeler, J. C. 2017, arXiv:1712.03240
\bibitem[Ruan et al. (2017)]{Ruan2017} Ruan, J. J., Nynka, M., Haggard, D., Kalogera, V. \& Evans, P. arXiv:1712.02809
\bibitem[Saxton et al.(2011)]{Saxton2011} Saxton, C. J., Barthelmy, S. D., D¡¯Elia, V., et al. 2011, GCN Circ. 12413
\bibitem[Sari et al.(1998)]{Sari1998} Sari, R., Piran. T., \& Narayan, R. 1998, \apj, 497, L117
\bibitem[Tanga et al. (2017)]{Tanga2017} Tanga, M.,  Kr\"{u}hler, T., Schady, P., Klose, S., Graham, J. F., Greiner, J., Kann, D. A., \& Nardini, M.
2017, arXiv:1708.06270
\bibitem[Tanaka \& Hotokezaka(2013)]{2013ApJ...775..113T} Tanaka, M., \& Hotokezaka, K.\ 2013, \apj, 775, 113
\bibitem[Tanvir et al.(2005)]{Tanvir2005} Tanvir, N. R., Chapman, R., Levan, A. J., Priddey, R. S. 2005, \nat, 438, 991
\bibitem[Tanvir et al.(2013)]{Tanvir2013} Tanvir, N.~R., Levan, A.~J., Fruchter, A.~S., et al.\ 2013, \nat, 500, 547
\bibitem[Tanvir et al.(2017)]{Tanvir2017} Tanvir, N.~R., Levan, A. J., Gonz\'{a}lez-Fern\'{a}ndez, C. et al. 2017, ApJL, 848, L27
\bibitem[Troja et al. (2008)]{Troja2008} Troja, E., Cummings, J. R.,  Palmer, D. M., Cucchiara, A., Barthelmy, S. D., Burrows, D. N.,
Roming, P., \& Gehrels, N. 2008, GCN Report 118.1
\bibitem[Troja et al. (2017)]{Troja2017} Troja, E., Piro, L., van Eerten, H. et al. 2017, Natur., 551, 71
\bibitem[Tsutsui et al.(2013)]{Tsutsui2013} Tsutsui, R., Yonetoku, D., Nakamura, T., Takahashi, K., \& Morihara, Y.\ 2013, \mnras, 431, 1398
\bibitem[Yang et al.(2015)]{2015NatCo...6E7323Y} Yang, B., Jin, Z.-P., Li, X., et al.\ 2015, Nature Communications, 6, 7323
%\bibitem[von Kienlin et al. (2017)]{von Kienlin2017}von Kienlin, A., et al., 2017,
%http://gcn.gsfc.nasa.gov/gcn3/21520.gcn3
\bibitem[Wanderman \& Piran (2015)]{Wanderman2015} Wanderman, D., \& Piran, T. 2015, MNRAS, 448, 3026
\bibitem[Wang et al.(2017a)]{WangH2017} Wang, H., et al.\ 2017a, ApJL, 851, L18 (arXiv:1710.05805)
\bibitem[Wang et al.(2017b)]{WangYZ2017} Wang, Y. Z., et al. \ 2017b, ApJL, 851, L20 (arXiv:1710.04781)
\bibitem[Wei \& Jin(2003)]{Wei2003} Wei, D.~M., \& Jin, Z.~P.\ 2003, \aap, 400, 415
\bibitem[Yamazaki et al.(2002)]{Yamazaki2002} Yamazaki, R., Ioka, K., \& Nakamura, T. 2002, \apjl, 571, L31
\bibitem[Zhang et al.(2012)]{ZhangFW2012} Zhang, F. W. et al.\  2012, ApJ, 750, 88\\

\end{thebibliography}
\end{document}